\documentclass[aps,prb,twocolumn,showpacs,superscriptaddress,groupedaddress]{revtex4}
\usepackage{graphicx}
\usepackage{amssymb}
\usepackage{amsmath}
\usepackage{bm}
\usepackage{color}
\usepackage{paralist}

\begin{document}

\title{Incoherent excitation and switching of spin states in exciton-polariton condensates}

\author{G. Li}
\affiliation{Nonlinear Physics Centre, Research School of Physics and Engineering, The Australian National University, Canberra ACT 0200, Australia}
\author{T. C. H. Liew}
\affiliation{Division of Physics and Applied Physics, Nanyang Technological University 637371, Singapore}
\author{O. A. Egorov}
\affiliation{Institute of Condensed Matter Theory and Solid State Optics, Abbe Center of Photonics, Friedrich-Schiller-Universit\"at Jena, Max-Wien-Platz
1, 07743 Jena, Germany}
\author{E. A. Ostrovskaya}
\affiliation{Nonlinear Physics Centre, Research School of Physics and Engineering, The Australian National University, Canberra ACT 0200, Australia}

%\date{\today}

\begin{abstract}
We investigate, theoretically and numerically, the spin dynamics of a two-component exciton-polariton condensate created and sustained by non-resonant spin-polarized optical pumping of a semiconductor microcavity. Using the open-dissipative mean-field model, we show that the existence of well defined phase-locked steady states of the condensate may lead to efficient switching and control of spin (polarization) states with a non-resonant excitation. Spatially inhomogeneous pulsed excitations can cause symmetry breaking in the pseudo-spin structure of the condensate and lead to formation of non-trivial spin textures. Our model is universally applicable to two weakly coupled polariton condensates, and therefore can also describe the behaviour of condensate populations and phases in 'double-well' type potentials. 
\end{abstract}

\maketitle

\section{Introduction}

Created in semiconductor microcavities, the two-component exciton-polariton condensates \cite{Deng10,Ciuti13} provide a rich playground for exploring the effects of  internal (pseudo-spin) degrees of freedom on the dynamics of non-equilibrium superfluids. A remarkable range of spin-related phenomena has been explored in systems with a coherent resonant optical pump, which offers the possibility to directly inject the spin state of exciton-polaritons and explore the ultrafast spin-switching dynamics \cite{RuboPRL10,Nmat10,Cerna13,Gavrilov13}. On the contrary, the non-resonant incoherent excitation responsible for the spontaneous formation of coherent exciton-polariton condensates is only beginning to be experimentally explored in the context of spin-dependent effects. The reason for this disparity is a long-held belief that the incoherent reservoir (uncondensed hot exciton-polaritons), inevitably created by the incoherent excitations far above the energy of the condensed exciton-polariton quasiparticles, lacks spin-selectivity, {\em i.e.}, owing to the fast spin-relaxation processes within the reservoir, the pump polarization cannot affect the spin of the condensed polaritons. Thus, the theoretical modelling of an incoherently excited exciton-polariton system with spin degrees of freedom usually treats the reservoir as either fast~\cite{RuboPRL10} or slow~\cite{BerloffPRB10} response scalar density state, without any spin discrimination.

Recent experimental studies of the optical spin-Hall effect \cite{NLSHO} employed microcavities with relatively short exciton lifetimes and succeeded in creating a spin-polarized reservoir of high-energy 'excitonic' polaritons. Such a reservoir can be controllably replenished by a pump laser of the selected polarization, thus opening up the way to create and manipulate spinor condensates of exciton-polaritons in the regime of incoherent far off-resonant optical excitation \cite{Anton14}.

In this work, we consider the exciton-polariton condensate in the regime of {\em cw} far off-resonant incoherent excitation, in the absence of external magnetic field. We analyse the non-stationary dynamics of condensate pseudo-spin and the relaxation dynamics towards possible steady states for both spatially homogeneous and inhomogeneous {\em cw} pump. We show that the spin-polarized incoherent reservoir allows efficient switching between different spin states of the polariton condensate, both in the spatially homogeneous and some inhomogeneous pumping regimes. Furthermore, we analyse steady-state spin textures and demonstrate that excitation of steady states by a spatially inhomogeneous pulsed optical pump can lead to the creation of stable non-trivial spin textures.

\section{The Model}\label{sec:the model}

We model the spontaneously created polariton condensate by the open-dissipative Gross-Pitaevskii equations \cite{Wouters07} derived from the Hamiltonian for a quantum fluid with pseudo-spin degrees of freedom obtained, in different contexts, in Refs. \cite{LiewPRB09,BerloffPRB10, RuboPRL10,Nmat10}. The dynamical equations for the condensate wavefunctions corresponding to the left $\psi_-$ and right $\psi_+$ hand circular polarization states are written as follows ($\sigma=\pm$) \cite{NLSHO}:
\begin{widetext}
\begin{equation}
i\hbar\frac{\partial \psi_{\sigma}}{\partial t}=\\
 \left(-\frac{\hbar^{2}}{2m}\nabla^{2}+u_a \lvert \psi_{\sigma} \rvert^{2}+u_b \lvert \psi_{-\sigma} \rvert^{2}+g_{R} n_\sigma
+\frac{i\hbar}{2}\left[Rn_{\sigma}(r)-\gamma_{c} \right] \right) \psi_{\sigma}+J \psi_{-\sigma}.
\label{model1}
\end{equation}
\end{widetext}
Here $m$ is the effective mass of the lower polarition, $\gamma_{c}$ is the
loss rate of polaritons which is the inverse of the polarition life
time, and $R$ characterises the stimulated scattering rate from the reservoir into the condensate. The interaction constants $u_a$ and $u_b$ characterise the scattering between the polaritons of the same and different polarization (spin) states, respectively. It is well established that  $|u_b|<|u_a|$ \cite{Vladimirova10}. Also, $g_R$ characterises   interactions between the reservoir and condensate, where for simplicity, we assume that interactions between oppositely spin polarized reservoir and condensed polaritons are negligible. The linear spin-mixing term proportional to $J$ (Josephson coupling) accounts for the sample-specific linear polarization splitting observed in experiments \cite{RuboPRL10,Nmat10} and usually arising due to stress at the semiconductor interfaces.

The equations for polariton wave functions are coupled to the equations for the reservoirs created by the incoherent {\it cw} pump with intensity $P_\sigma (r,t)$. It is usually assumed that  all memory of the polarization of the incoherent pump is lost \cite{LiewPRB09,RuboPRL10}. However, as mentioned above, it was shown that the creation of a spin-polarized exciton reservoir is possible with a spin-imprinting optical pump for sufficiently short-lived excitons. Because the precise spin-relaxation mechanisms for the reservoir are unknown, we will consider the simplest model of spin-dependent reservoir, which was introduced in  \cite{NLSHO}: 
\begin{equation}\label{eq:bare n}
\frac{\partial n_\sigma}{\partial t}=P_\sigma(r,t)-\left(\gamma_R+R \lvert \psi_\sigma \rvert^2\right)n_{\sigma},
\end{equation}
where $\gamma_R$ is the loss rate of
the reservoir polaritons and $P_\sigma(r,t)$ is the pump.
The dynamics of the reservoir is assumed to be fast on the scale of the polariton lifetime, \emph{i.e.}, comparable to the dynamics of the condensate, so that the model of slow-response, static reservoir \cite{BerloffPRB10} is no longer applicable.  We note that the coherent resonant injection of spin  \cite{Nmat10} can be readily introduced into these equations.

By introducing the characteristic time $T=\gamma_{c}^{-1}$, length  $L=\sqrt{\hbar/(m\gamma_{c})}$, and energy $E_{u}=\hbar \gamma_c$ scales \cite{parameters}, we can rewrite Eq.~(\ref{model1}) and (\ref{eq:bare n}) in the dimensionless form:
%\begin{widetext}
\begin{eqnarray}\label{eq:rescaled}
\begin{aligned}
i\partial_{t}\psi_{\sigma} &=\left\{ -\frac{1}{2}\nabla^{2}+u_{a} \lvert \psi_{\sigma} \rvert^{2}+u_{b} \lvert \psi_{-\sigma} \rvert^{2}+g_R n_\sigma \right.\\
                           &\qquad \left.+\frac{i}{2}\left[Rn_{\sigma}-\gamma_c\right]\right\} \psi_{\sigma}+J\psi_{-\sigma} \\ \label{rescaled}
\partial_{t}n_{\sigma}&=P_\sigma(r)-(\gamma_R+R \lvert \psi_\sigma \rvert^2)n_\sigma\label{rescaled_nR}\\  
\end{aligned}
\end{eqnarray}
%\end{widetext}
where $u_{a}\to u_a/(E_{u} L^2)$, $u_{b}\to u_b/(E_{u} L^2)$, $g_R\to g_R/(E_{u} L^2)$, and $J\to J/E_{u}$. In these scaling units, $\gamma_c=1$, however, we will retain it in the subsequent formulas for clarity. In the following discussions, except some specified cases, all parameters take the numerical values indicated in Ref. \cite{parameters}.

%We note that, for moderate values of the pump intensity and in the absence of linear coupling $J=0$, these parameters place the system into the diffusive (overdamped) regime of spatially homogeneous density perturbations, which is defined by the condition $\gamma_c/\gamma_R<{\bar P_\pm}^2/4({\bar P_\pm}-1)$, where ${\bar P_\pm}$ is the pump intensity normalised by the threshold value, $P_{th}=\gamma_c\gamma_R/R$ \cite{Wouters07}. This implies that a spatially narrow excitation of a stationary state, created by a spin-polarized pulsed pump, will have significant fraction of supersonic group velocity components and therefore expected to produce dispersive shock wave behaviour, such as generation of dark soliton stripes and vortices at large excitation densities.

\section{Homogeneous Steady States}\label{Homogeneous Steady States}

We start with the analysis of the spin-mixing dynamics for a spatially homogeneous condensate. In experiments, the quasi-homogeneous distribution of condensate density can be achieved, {\em e.g.}, by a flat-top 'super-Gaussian' excitation. In this case, it is convenient to separate the amplitude and phase dynamics of the order parameter by using the transformation: $\psi_\sigma(t)=\Phi_\sigma(t)\exp{(i\phi_\sigma(t))}$ and introduce the density of the spinor components $\rho_\sigma(t)\equiv \Phi^2_\sigma$ and the relative phase difference $\theta=\phi_{-}-\phi_{+}$. Substituting these spatially homogeneous states into Eq. (\ref{eq:rescaled}), we obtain the dynamical equations for the condensate density $\rho_\sigma$, the relative phase $\theta$, and the reservoir density $n_\sigma$:
\begin{equation}\label{eq:homo_dynamics}
\begin{aligned}
\dot{\rho}_{\sigma}&=(Rn_\sigma-\gamma_{c})\rho_{\sigma}+\sigma 2J\sqrt{\rho_{\sigma}\rho_{-\sigma}}\sin\theta \\
\dot{\theta}\;\:& = \left(U-\frac{J\cos\theta}{\sqrt{\rho_+ \rho_-}}\right)(\rho_{+}-\rho_{-})+g_R \delta n_R\\
								% \frac{\cos\theta}{\Phi_{+}\Phi_{-}}\right](\Phi_{+}^{2}-\Phi_{-}^{2})\\
\dot{n}_\sigma&=P_\sigma-(\gamma_R+R \rho_\sigma) n_\sigma,\\             %!!!
\end{aligned}
\end{equation}
where $U=u_a-u_b$ and $\delta n_R=n_+-n_-$. The dynamics described by Eq. (\ref{eq:homo_dynamics}) is rather universal, and would apply to the description of any two-mode polariton system with linear (Josephson-type) coupling such as condensates in a double-well potential \cite{Lagoudakis10} or a two-level system \cite{Eastham08}, provided that their spatial variation can be ignored.
%
%The system (\ref{eq:homo_dynamics}) is completely equivalent to Eq. (\ref{eq:rescaled}), and its dynamics describes evolution of the partial condensate densities in the two spin states. 

The amplitude and phase of each component in the condensate determine the polarization of the coherent photoluminescence emitted from the cavity. %, and therefore the pseudo-spin state of the polariton uniquely defines the polarization state of light. 
 It is therefore helpful to use the pseudo-spin representation (see, {\em e.g.}, \cite{LiewPRB09}) which allows us to visualise the polarization of the condensate as a point on the Poincar\'{e} (Bloch) sphere. 
The coordinates of the point are given by the components of the Stokes vector defined in the standard manner:
\begin{equation} \label{eq:sxyz}
\begin{aligned}
S_x&=\sqrt{\rho_{\sigma}\rho_{-\sigma}}\cos\theta,\,S_y=\sqrt{\rho_{\sigma}\rho_{-\sigma}}\sin\theta,\, \\ S_z&=\frac{1}{2}(\rho_{+}-\rho_{-}),\, S_0=\frac{1}{2}(\rho_{+}+\rho_{-})
\end{aligned}
\end{equation}
with the total condensate density $n_c=\rho_++\rho_-$ defining the length of the Stokes vector: $S^2_0=S^2_x+S^2_y+S^2_z=n^2_c/4$. 
%The dynamics of the Stokes vector obeys the following equations:
%\begin{equation}\label{eq:stokes}
%\begin{aligned}
%\dot{S_x}&=\frac{1}{2}\left(R n_R-2\gamma_c\right) S_x-2US_yS_z-g_R\delta n_R S_y,\\
%\dot{S_y}&=\frac{1}{2}\left(R n_R-2 \gamma_c\right) S_y+2US_xS_z+g_R\delta n_R S_x-JS_z,\\
%\dot{S_z}&=\frac{R}{2}\left(S_0 \delta n_R+S_z n_R\right)-\frac{\gamma_c}{2}S_z+2JS_y,\\
%\dot{S_0}&=\frac{R}{2}\left(S_0 n_R+S_z\delta n_R\right)-\frac{\gamma_c}{2}S_0,
%\end{aligned}
%\end{equation}
%where we have introduced the notation $n_R=n_{+} + n_{-}$.
Note that both the pseudo-spin and the total condensate density, $S_0=n_c /2$, are functions of time, therefore it makes sense to consider the evolution of this vector in the normalized pseudo-spin space, $s_i(t)=S_i(t)/S_0 (t)$, {\em i.e.}, on the Poincar\'{e} sphere of unit radius. Although we will use the normalized Poincar\'{e}  sphere representation throughout the text, one should keep it in mind that its (not normalized) radius can change dynamically.

The stationary states of the system (\ref{eq:homo_dynamics}) define fixed points on the Poincar\'{e} sphere, which correspond to synchronized, phase-locked solutions of the Eq. (\ref{eq:rescaled}) with a well-defined polarization state of the polaritons. Desynchronised states with a time-dependent phase correspond to Josephson oscillations, and, in the case of steady undamped oscillations Ð to limit cycles on the Poincar\'{e} sphere. 

In principle, steady states can be obtained in a closed analytical form, however the expressions are cumbersome, and a more useful insight into the system's dynamics is obtained in some limiting cases. As noted in \cite{Wouters08,BerloffPRB10}, the following conditions are of particular interest: (a) the excitons decay into the polariton channel dominates the relaxation dynamics of the reservoir, $R\rho_\pm\gg \gamma_R$; (b) the polariton interaction energy is greater than the tunnelling energy, $Un_c\gg J$, which  approximately defines the boundary of the synchronization region; (c) populations of the two spin components are similar (semi-linearly polarized pumping), $n_c\gg S_z$.
 Following the procedure outlined in \cite{BerloffPRB10}, one can derive the approximate self-consistent equation for the relative phase, which describes a Josephson junction driven by a constant bias current (see \cite{Steven01}, Ch. 8.5):
\begin{equation}\label{theta}
\frac{1}{2}\,\ddot{\theta}+\frac{\gamma_c}{2}\,\dot{\theta}= (U+g_R)\frac{\delta P}{2}+UJn^0_c\sin \theta,
\end{equation}
here $\delta P=P_+-P_-$, $n^0_c=(P_++P_-)/\gamma_c$. Analysis of the phase space of this dynamical system shows that, for the equal pumping of the two spin components, i.e. linearly polarized pump, $\delta P =0$, only phase-locked states exist. The system becomes  desynchronised for an elliptically polarized pump when $\lvert I_c \rvert=\left[\lvert \delta P \rvert /(P_++P_-)\right] \left[\gamma_c(U+g_R)/(2UJ)\right]>1$, whereby the dynamics is dominated by limit cycles. For $\lvert I_c \rvert<1$ the system admits fixed points  determined by the condition $\sin \theta=I_c$ and corresponding to the distinct steady states of the homogeneous system for each value of the pump intensity. 

Although Eq.~(\ref{theta}) is obtained under a very restrictive assumption of a nearly linear polarization of the polaritonic state, it can describe some experimental results such as in Ref. \cite{Lagoudakis10}. Notably, when the pump power is large, conditions (a) and (b) can be automatically fulfilled once condition (c) is satisfied. Physically the condition (c) means that the system effectively evolves on a fixed-radius Poincar\'{e}  sphere,  where its evolution is then given by the competition between the nonlinear polarition-polarition interaction and the Josephson tunnelling between two components. Meanwhile, if the pump power is large, then the polarition-polariton interaction dominates and Eq. (\ref{theta}) is applicable (detailed discussion can be found in Ref.~\cite{BerloffPRB10}). 

%Moreover, it is derived under the assumption of a steady-state reservoir, which may not be justified for our model accounting for a fast-responding excitonic reservoir.
In general, however, Eq.~(\ref{theta}) does not always apply, especially when the Josephson coupling dominates. Therefore, in what follows we will consider steady states of Eq. (\ref{eq:homo_dynamics}) given by the pumping conditions of different polarzation and strength. 

%And we will identify a scheme for controlled switching between different polarization states of the polaritons and subsequently the photoluminescence signal. 

\subsection{Linearly polarized pump}\label{subsec:Linearly polarized pump}

{\begin{figure}[t]
\includegraphics[width=\columnwidth ]{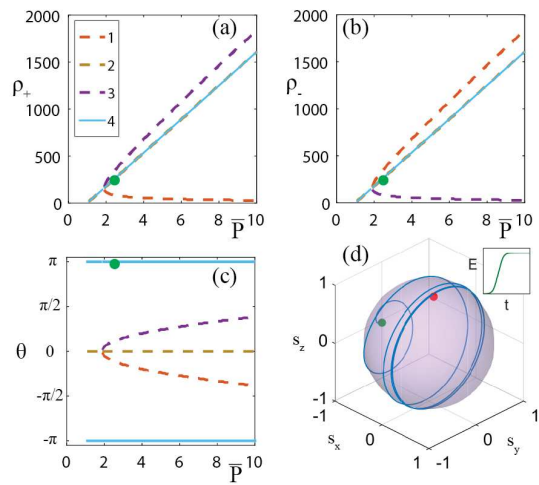}
\caption{(Color online). (a)-(c) Fixed point solutions of Eq.~(\ref{eq:homo_dynamics}) under balanced (linearly polarized) pumping.
Branch $1$ and $3$ are elliptically polarized solutions; branch $2$ and $4$ are bonding ($s_1$) and anti-bonding ($s_2$) solutions respectively.
Only the solid branch is sable (see text). Green dots: corresponding points to the inhomogeneous simulation. (d) (see Sec.~\ref{sec:inhomogeneous}) Inhomogeneous simulation results of Eq.~(\ref{eq:rescaled}). Evolution of the integrated Stokes parameters in phase space under balanced Gaussian-shape pumps. Parameters (see Sec.~\ref{sec:inhomogeneous} for the definitions): $\bar{P}=5, P^L_{im}=0.5, a_x=a_y=5$, \text{and} J=0.5. Red dot: initial state of evolution; green dot: final state of evolution. Inset: energy evolution.}
\label{solution_eqPump}
\end{figure}

Most commonly, the optical pump producing a polariton condensate in non-spin-resolved experiments is linearly polarized. In our model, this means that the pump is balanced: $P_+=P_-$. Under this condition, from the first of Eq. (\ref{eq:homo_dynamics}), it follows that the synchronized state can be reached when $Rn_\sigma=\gamma_c$ and the internal Josephson current $I_J=2J\sqrt{\rho_+ \rho_- }\sin\theta$ becomes zero. Consequently, the phases are locked to $\theta = 0$ or $\theta = \pi$ (mod $2\pi$). The two fixed points on the Poincar\'{e}  sphere corresponding to these phase locked linearly polarized states states are $s_1 = (1, 0, 0)$ (corresponding to $\theta = 0$), and $s_2=(-1, 0, 0)$ (corresponding to $\theta = \pi$). The density of the steady state grows linearly with the pump rate: $\rho_\sigma=(P_{th}/\gamma_c)\left(P_{\sigma}/P_{th}-1\right)$, see Fig.~\ref{solution_eqPump} (a-c), where the normalized pump power is defined as $\bar{P}_\sigma=P_\sigma / P_{th}$ and for the linearly polarized pump particularly $\bar{P}_+=\bar{P}_-=\bar{P}$.

Following the standard Lyapunov stability analysis \cite{Steven01}, we deduce that, for our choice of the sign of $J$, the $\pi$ out-of-phase state $s_2$ (the so called \emph{anti-bonding state}) is stable and the in-phase state $s_1$ is unstable. Direct numerical integration of Eq. (\ref{eq:homo_dynamics}) with different initial conditions also confirm that $s_2$ is stable Fig.~\ref{Bloch_sphere}(a-b). We note that this selection of the anti-bonding steady state in a weakly (linearly) coupled two-state system has also been recently confirmed in \cite{Lien14,Tosi14}. This effect persists even for the inhomogeneous (Gaussian) pumping \cite{Tosi14}, as shown in Fig. \ref{solution_eqPump}(d) and discussed in Sec.  ~\ref{sec:inhomogeneous} below. 

The other steady states arise when $\rho_\sigma>(P_{th}/\gamma_c)\left(P_{\sigma}/P_{th}-1\right)$, where the sign of the inequality is the opposite for the other polarization component, see Fig.~\ref{solution_eqPump} (a-c) dashed branch $1$ and $3$. 
They are sustained by a non-zero internal Josephson current which will grow with the pumping power. Lyapunov analysis reveals that they are unstable. 

\begin{figure}
\includegraphics[width=8.5cm]{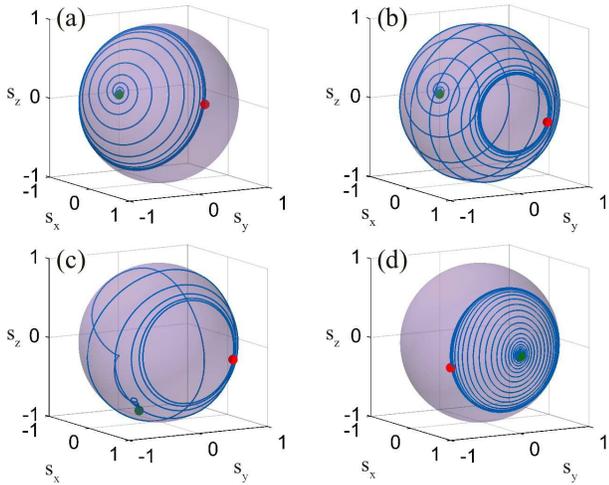} 
\caption{(Color online). Trajectories of the pseudo-spin vector obtained from numerical solutions of Eq. (\ref{eq:homo_dynamics}) with $J=0.5$ and various interaction strengths between the condensate and reservoir $g_R$. Red dots: initial states of evolution; Green dots: final states of evolution. Initial conditions are arbitrary. 
Parameters are:
(a)$g_R=1.5\times 10^{-2}, \theta(0)=0.5$; (b)$g_R=1.5\times 10^{-2}, \theta(0)=2$; (c) Artificially enlarged $g_R=1.5\times 10^{-1}$;
(d) Artificially reduced $g_R=1.5\times 10^{-3}$.}
\label{Bloch_sphere}
\end{figure}

Introduced by the interaction between the condensate and the reservoir, the blue shift term which is proportional to $g_R$ plays an important role in the spin dynamics.
It can modify the fixed point properties when its energy is compatible with the Josephson tunnelling energy.
Starting with some arbitrarily chosen small amplitude initial conditions,
Fig.~\ref{Bloch_sphere}(c-d) illustrates how different values of $g_R$ can affect the evolution of Stokes parameters in phase space.
Indeed, both the positions of fixed points and their stability can change, e.g., when $g_R=1.5\times 10^{-3}$, both $s_2$ and $s_1$ become stable, see Fig.~\ref{Bloch_sphere}(d); higher value of $g_R$ will lead to the system in an elliptically polarized state, see Fig.~\ref{Bloch_sphere}(c). Here, the trajectories are obtained by numerical integration of Eqs. (\ref{eq:homo_dynamics}). 

\subsection{Elliptically polarized pump}\label{subsec:elliptically polarized pump}

\begin{figure}
\includegraphics[width=\columnwidth ]{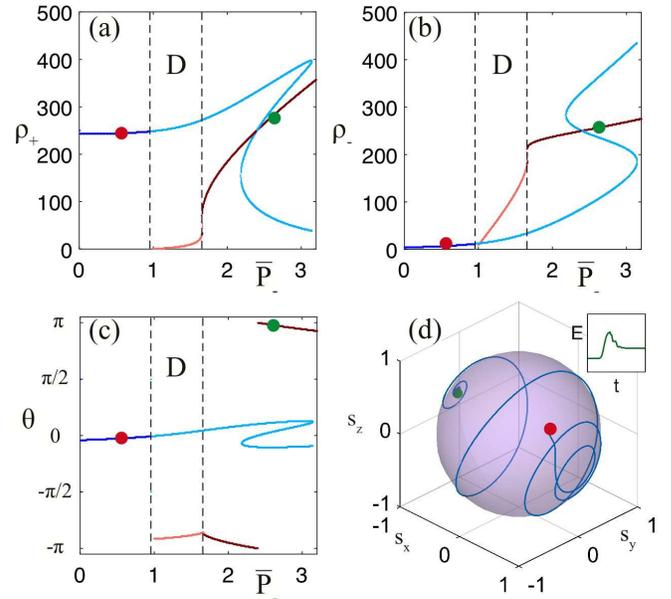}
\caption{(Color online). (a)-(c) Selected fixed point solutions of Eq.~(\ref{eq:homo_dynamics}), with an elliptically polarized pump for $\bar{P}_{+}=2.4$. Curves with darker color are stable while those with lighter color are unstable. Red (green) dot: the steady state before (after) a pulse is added.
(d)  Inhomogeneous simulation results of Eq.~(\ref{eq:rescaled}) (see Sec.~\ref{sec:inhomogeneous}). Evolution of integrated Stokes parameters in phase space. Pumping parameters: $\bar{P}=4, P^L_{im}=0.4, a_x=a_y=5, J=0.5$; Pulse parameters: $\bar{P}^{se}=0.12, P^L_{im}=1, a_{px}=a_{py}=5$. (see Sec.~\ref{sec:inhomogeneous} for the definitions.) Red dot: initial state of evolution; green dot: final state of evolution. Inset: energy evolution.}
\label{NoneqPump_Pminusvaries}
\end{figure}

When the imbalanced polarization-selective pumping is introduced, the dependence of the density of the orthogonal spin components on the pump power displays a hysteresis-like profile. However, it does not correspond to the typical bistability behaviour, since most of the (light blue)curve corresponds to an oscillatory unstable fixed point.
Fig.~\ref{NoneqPump_Pminusvaries} (a-c) shows a typical fixed point distribution against small pumping imbalance, where all unstable solution branches with at least one vanishing $\rho_\sigma$ are not shown for clarity. Stable fixed points exist in semi-circular polarized pumping region (dark blue), which will be discussed in the next subsection, and the semi-linearly polarized pumping region (dark red), where at the point $P_-=P_+$ the fixed point coincides with the single anti-bonding stable state $\rho_+=\rho_-$ shown in Fig. \ref{solution_eqPump}.

%albeit with very small growth rate, which enables the system to stay in the vicinity of the steady state for a long time. 
%Only one steady state is stable in this regime, which for  

One should note that the above discussion is based on the Lyapunov stability analysis, which describes the linear stability to a long-wavelength (spatially homogeneous) perturbation with the wave vector ${\bf k}=0$. The steady state may become unstable to a spatially modulated perturbation with ${\bf k} \neq 0$, see, {\em e.g.}, discussions about the {\em modulational instability} in 2D \cite{BerloffPRB10} and 1D \cite{Kamchatnov13} cases.

%and this  has been considered both in   spin-polarized polariton condensates.

%As a verification of each other, one should note that Fig. \ref{solution_eqPump} (a-c) and Fig. \ref{Bloch_sphere} (a,b) were obtained from different methods. Whereas the former was given by solving numerically the static Eq. (\ref{eq:homo_dynamics}) which then became nonlinear algebraic equations, the later was given by numerical integration of Eq. (\ref{eq:homo_dynamics}), starting with different initial conditions. They fit each other very well. Further, they also fit the inhomogeneous simulation results qualitatively, see 
%Fig. \ref{solution_eqPump} (d) and Sec.~\ref{sec:inhomogeneous} for detailed discussion.

As a verification of the existence of the stable branches, we performed a pulsed excitation of a steady state in Fig. \ref{NoneqPump_Pminusvaries} (a-c) where the red and green dots represent stable steady states before and after the pulse. They fit the inhomogeneous simulation result, Fig. \ref{NoneqPump_Pminusvaries} (d), qualitatively well. (see Sec.~\ref{sec:inhomogeneous} for details.)

\subsection{Circularly polarized pump}

The steady state shown in Fig. \ref{NoneqPump_Pminusvaries} (a-c) at $P_-=0$ corresponds to the situation when only one spin component is pumped, which is physically akin to the strongly imbalanced dissipative double-well Josephson junction \cite{Lien14}. (The opposite case where $P_+=0$ is similar.)
In this limiting case, the fixed point solutions of Eq. (\ref{eq:homo_dynamics}) are shown in Fig.~\ref{solution_Pplusvaries}, where the steady state for $\bar{P}_+=2.4$ is the same as that represented by the dark blue curve in Fig. \ref{NoneqPump_Pminusvaries} (a-c) at $P_-=0$.
 
The stable branch whose $\theta$ is small but never vanishes corresponds to the so called \emph{self-trapped state} with a strongly imbalanced population. It is a consequence of nonlinear interactions \cite{Leggett01,atom_MQST} and, in the context of a double-well polariton system, has been studied both experimentally \cite{Bloch_MQST} and theoretically \cite{Lien14}. 

Again, Fig.~\ref{solution_Pplusvaries} (d) shows us the inhomogeneous simulation result verifying the position of fixed points predicted by the homogeneous model.%{\em Where are the corresponding points in panels (a-c)?} }

\begin{figure}[t]
\includegraphics[width=\columnwidth ]{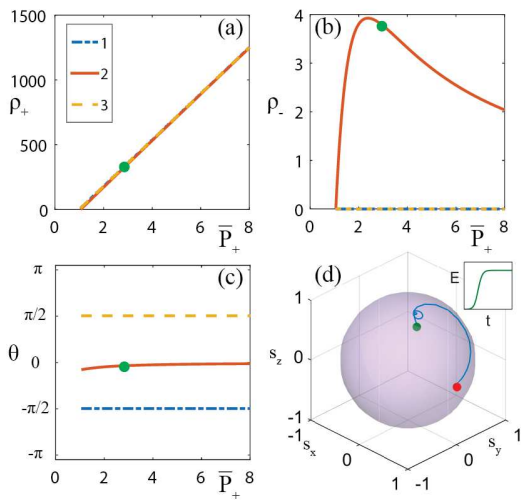}
\caption{(Color online). (a)-(c) Fixed point solutions of Eq.~(\ref{eq:homo_dynamics}), with $\bar{P}_{-}=0$. Only the solid branch is stable (see text). Green dots: corresponding points to the inhomogeneous simulation. (d) (see Sec.~\ref{sec:inhomogeneous}) Inhomogeneous simulation results of Eq.~(\ref{eq:rescaled}). Evolution of integrated Stokes parameters in phase space under circularly polarized pump. Parameters: $\bar{P}=3, P^L_{im}=0, a_x=a_y=5$, \text{and} J=0.5. Red dot: initial state of evolution; green dot: final state of evolution. Inset: energy evolution. (see Sec.~\ref{sec:inhomogeneous} for the definitions.)}
\label{solution_Pplusvaries}
\end{figure}

\section{Inhomogeneous spin-switching dynamics}\label{sec:inhomogeneous}

In the spatially inhomogeneous situation, it is well known that the in-plane magnetic fields, either being externally applied or caused by the TM-TE splitting \cite{Kavokin04}, can lead to nontrivial spatial structures for both density and spin distributions \cite{BerloffPRB10,Kavokin05,NLSHO}.
The linear spin coupling can also be regarded as an effective magnetic field pointing along the $x$ axis, which can lead to the appearance of spin patterns. However,  here we analyse steady states of the polariton spinor systems  sustained by a {\em cw} pump with a Gaussian shape and show that the {\em spatially averaged} evolution of the spin components qualitatively follows the prediction of the homogeneous model discussed in Sec. \ref{Homogeneous Steady States}. The results obtained within the homogeneous approximation can therefore be used as a guide for spin-switching manipulation with a realistic, spatially inhomogeneous pump.

%{\color{red} Please check the notations in the paragraph below and Fig. \ref{round_pump_steady_state} (and others) as they were/are most confusing. Why cannot you simply assume that $P_0 =P_-+P_+$? Also, if $P$ depends on $(x,y)$, what is the number corresponding to $\bar{P}$ in figure captions? Peak pump? In which case, should it be $\bar{P}_0$?}
To determine the evolution of pseudo-spin under spatially inhomogeneous excitation, we solve directly
the open-dissipative GP equations (\ref{eq:rescaled}) with a radially symmetric Gaussian pump by using the split-step method \cite{Javan06}.
The total intensity of the pump is defined as $P(x,y)=P_0 e^{-(x^2/a^2_x+y^2/a^2_y)}$, where $a_{x,y}$ is the width along the $x$ and the $y$ direction.
The pump for each circularly polarized component is given by splitting the energy from $P(x,y)$, which is $P(x,y)=P_+(x,y) + P_-(x,y)$,
since in many experiments the pump is given by a single laser~\cite{NLSHO}. %, {\em e.g.}, \cite{NLSHO}.
It is convenient to use the spin bias defined as $P^L_{im}=P_0^- / P_0$ to denote those two pumps, 
then $P_-(x,y)= P^L_{im} P(x,y)$ and $P_+(x,y)= (1-P^L_{im}) P(x,y)$.
Also, throughout the text, the notation of the normalized (inhomogeneous) pump is given by,
 $\bar{P}=P_0/P_{th}$ (normalization against its peak value), where the pump threshold $P_{th}$ is the same as that in Sec.\ref{Homogeneous Steady States}.

By adjusting the pumping power and the imbalance value, with an arbitrary small amplitude initial condition,
the system can be driven to a synchronized steady state without any external potential.
The condensate mean-field energy functional, $E=E_+ + E_-$, serving as the measure of kinetic and interaction energy of the condensate  \cite{Bao13}, can be used to detect whether the steady state is reached:
\begin{equation}\label{energy}
\begin{aligned}
E_{\sigma}&=\int d{\bm r}[ \lvert \nabla \psi_\sigma \rvert^2/2+u_a \lvert \psi_\sigma \rvert^4+g_R n_\sigma \lvert \psi_\sigma \rvert^2+\\
&+u_b \lvert \psi_\sigma \rvert^2 \lvert \psi_{-\sigma} \rvert^2+J\text{Re}(\psi^*_\sigma\psi_\sigma)], 
\end{aligned}
\end{equation}
where the integration is over the area where the condensate density is non-negligible.

To map the evolution of the inhomogeneous condensate to that of the Stokes parameters, we trace the evolution of the spatially integrated quantities as  $s^{\rm int}_i(t)=\int S_i (x,y) d{\bm r}/S_0^{int}$, where $S_0^{int}$ is given by the spatially integrated length of the Stokes vector. The integration area is the same as that used to determine the steady state energy (\ref{energy}).  In the following, unless it is indicated explicitly; we shall refer to the normalized $s_{i}^{int}(t)$ simply as integrated Stokes parameters. This normalization procedure, although eliminating information on the spatial distribution of the pseudo-spin, allows us to recover the spatially averaged polarization state. This averaged spin dynamics could be observable experimentally even without performing spatially resolved polarimetry of the cavity photoluminescence. 

\begin{figure}[t]
\includegraphics[width=8.5cm]{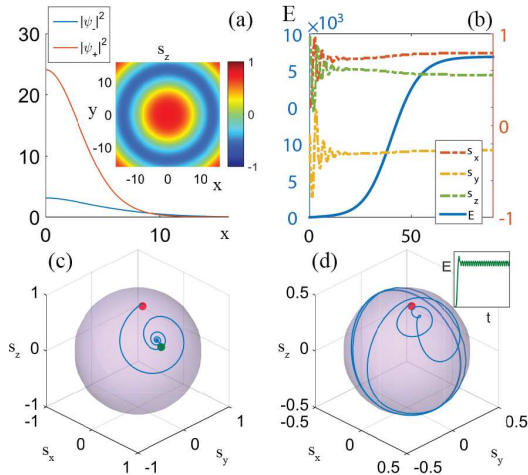}
\caption{(Color online). (a)-(c) Synchronized steady state, with $\bar{P}=4, P^L_{im}=0.4,a_x=a_y=5$, and $J=0.5$: (a) Density profiles (along $y=0$) and spatial distribution of $s_z$ for the steady state. (b) Evolution of integrated Stokes parameters and energy. (c) Evolution of integrated Stokes parameters in phase space. Starting form the red dot; ending with the green dot. (d) desynchronised state, with $\bar{P}=8,  P^L_{im}=0.4, a_x=a_y=5$, and $J=0.5$: evolution of integrated Stokes parameters. Starting form the red dot; ending with a closed orbit.% Here $\bar{P}=P / P_{th}$
Inset: energy evolution. The integrations were over areas where $n_c(x,\, y)>10^{-3}$.}
\label{round_pump_steady_state}
\end{figure}

Polarization dynamics for the linearly  and circularly polarized pump are shown in Fig. \ref{solution_eqPump}(d) and Fig. \ref{solution_Pplusvaries}(d), respectively. As we can see, the spatially averaged dynamics in the case of inhomogeneous pumping fits the homogeneous predictions well.

For a typical steady state under elliptically polarized  pumping, the value of the cross section densities and spatial distribution of the pseudo-spin is shown in Fig.~\ref{round_pump_steady_state}(a).
Its build up process is similar to that of the linearly-polarized pumping case, see Fig.~\ref{round_pump_steady_state}(b-c). 
The linear coupling $J$ causes switching of the dominant density between the two components as polaritons spread out from the pump source, which leads to the appearance of radially symmetric domains of polarization density $s_z$. The larger $J$ leads to the denser spin pattern and loss of correspondence between the averaged spin dynamics in inhomogeneous and homogeneous cases. If $J$ is sufficiently small, the polarization state is almost homogeneous across most of the integration area. 

\begin{figure}[t]
\includegraphics[width=\columnwidth ]{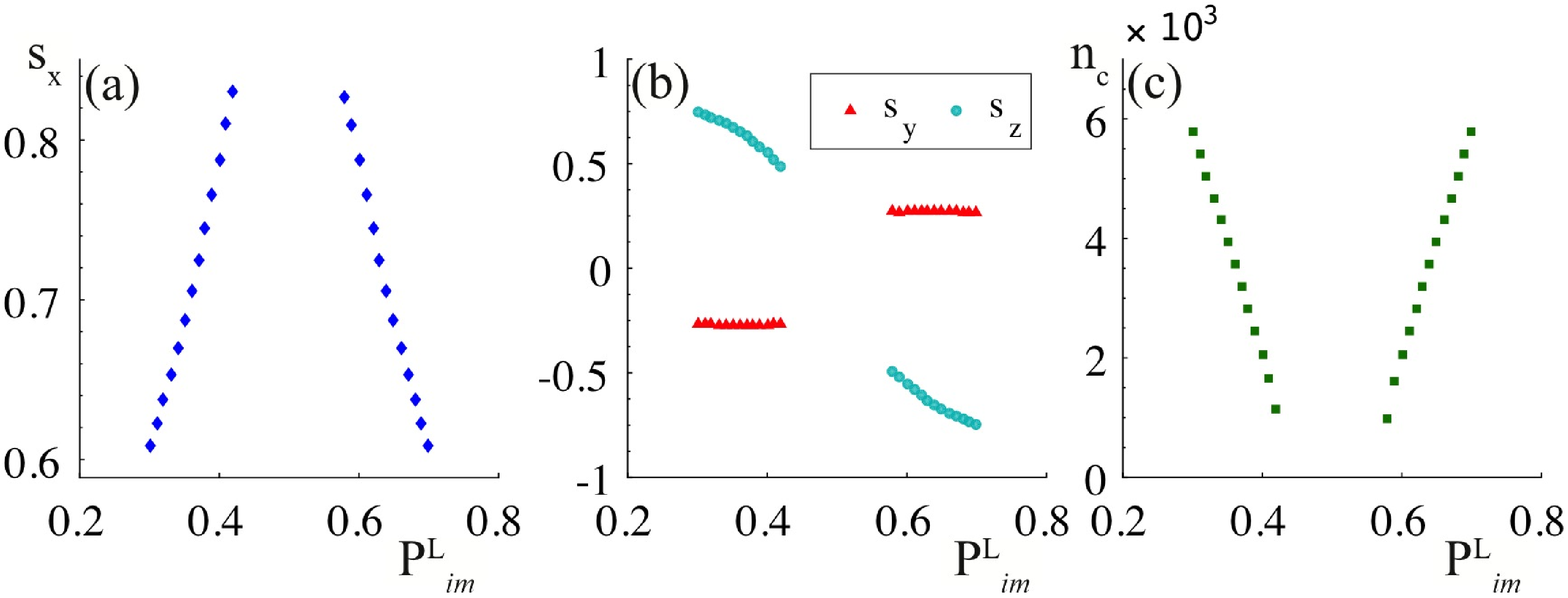}
\caption{(Color online). Steady state integrated quantities against the pump imbalance $P^L_{im}$. Parameters: $\bar{P}=4, a_x=a_y=5, J=0.5$. Between the gap $0.42<P^L_{im}<0.58$, one needs to increase the $\bar{P}$ value to reach steady states (see text).}
\label{scatter}
\end{figure}

The degree of control of the averaged polarization state of the condensate attainable by an incoherent, far off-resonant spin-polarized excitation according to our model is presented in Fig.~\ref{scatter} and could be tested in an experiment, which would validate our assumptions.
Due to our choice of the pump power distribution configuration,
in Fig.~\ref{scatter}, between $0.42<P^L_{im}<0.58$ one needs to increase $\bar{P}$ in order to reach steady states. 
With a larger pump power, however, the degree of control of polarization decreases because  
the system will easily fall into a desynchronised state even with small value of imbalance.
% which will can be understood as the area of region 'D' (see Fig. \ref{NoneqPump_Pminusvaries}) increases with the pump power.

For example, for an elliptically polarised pump, within the region marked with 'D' (which stands for desynchronised) in Fig. \ref{NoneqPump_Pminusvaries}, none of the fixed point branches are stable, which does not rule out the existence of the closed orbits (limited cycles) on the Poincare sphere corresponding to desynchronised states.
As shown in Ref. \cite{Wouters08}, for a desynchronised state the density $\rho_\sigma (t)$ maintains periodic oscillations and the evolution of its Stokes vector is similar to Fig. \ref{round_pump_steady_state} (d) where the angle of the trajectory plane can vary depending on the system parameters. 
This regime cannot be accessed by Eq.~(\ref{theta}) because the latter can only describe limit cycles in the vicinity of the $s_z=0$ plane.
Indeed, it can be revealed by inhomogeneous simulations of the full model equations that,
in such a state, the spatial distribution of the spin pattern keeps breathing.
The corresponding Stokes parameters circle around the Poincar\'{e} sphere \cite{Gavrilov14}, reducing the degree of (time-averaged) polarization of its luminescence. A similar pinning and depinning effect has been observed experimentally \cite{RuboPRL10}.  Here this effect is associated with moving in and out of the phase-locked synchronized regime of the condensate dynamics caused by intrinsic interactions between the two spin components, so that it can arise even without taking into account the structural disorder in the sample.

\begin{figure}[t]
\includegraphics[width=\columnwidth ]{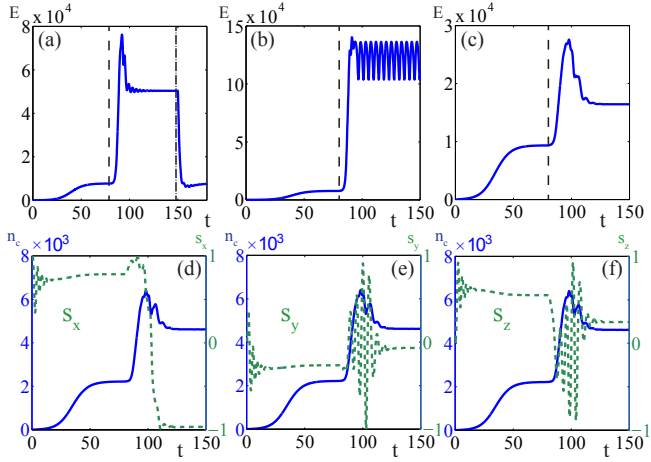}
\caption{(Color online). Spin switching dynamics for the case of Gaussian pump. (a-c) Energy of the initial and final state after the switching pulse. Dash line: pulse switched on. Dash-dot line: pulse switched off; (d-f) Integrated total density and integrated Stokes parameters during the switching process for switching between the steady states shown in (c). Parameters: the same as Fig.~\ref{NoneqPump_Pminusvaries} (d), except for (a) $\bar{P}^{se}=0.4, P^L_{im}=0.65$; and (b) $\bar{P}^{se}=0.9, P^L_{im}=0.65$.}
\label{switch_inh}
\end{figure}

%Pumping parameters: $\bar{P}=4, P^L_{im}=0.4, a_x=a_y=5, J=0.5$; Pulse parameters: $\bar{P}^{se}=0.12, P^L_{im}=1, a_{px}=a_{py}=5$. (see Sec.~\ref{sec:inhomogeneous} for the definitions.) Red dot: initial state of evolution; green dot: final state of evolution. Inset: energy evolution.

The coexistence of stable and unstable fixed points described in Sec.~\ref{Homogeneous Steady States} indicates that an inhomogeneous pump might be used to switch the system between different density and polarization states. Indeed, we find that the efficient switching between different spin and density states can be achieved by applying a radially symmetric incoherent {\em pulsed excitation} of the form: $P_{se}({\bf r},t)=(1/4) P_0^{se}\{ 1+\tanh[\tau_p (t-t_0)]\}\{ 1+\tanh[\tau_p (t_1-t)] \} e^{-(x^2/a^2_{px}+y^2/a^2_{py})}$, where $P_0^{se}$ is the peak value, $a_{px,py}$ determines the pulse width, $t_0$ and $t_1$ ($t_1>t_0$) are the pulse switch on and switch off time respectively, and $\tau_p$ is a coefficient controlling adiabaticity of the excitation (large $\tau_p$ corresponds to a non-adiabatic pulse).
The normalized notation for the pulse is given by $\bar{P}^{se}=P_0^{se}/P_{th}$.
Note that the combined intensity after the pulse was added is given by the interference between pump and pulse modes.
Unless specified, they are regarded as being in phase. 
A radially symmetric non-adiabatic pulse,  $a_{px}=a_{py}$, causes an abrupt change in the condensate energy and a strong outward density flows combined with the internal Josephson currents leading to spin mixing dynamics [Fig. \ref{switch_inh}(d-f)]. As a result, the system can enter a different steady state [Fig. \ref{switch_inh}(a,c)], or a phase desynchronised state [Fig. \ref{switch_inh}(b)]. If the original steady state is stable, the perturbed system restores its initial density and polarization state after the pulse is switched off.

Regardless of the fact that with strong spatial variation the homogeneous results would not be generally applicable, one can still compare the spatially-averaged inhomogeneous results of Fig. \ref{switch_inh}(c-f) with the fixed point solutions of Eq. (\ref{eq:homo_dynamics}), as long as the former begins and ends with steady states. Fig. \ref{NoneqPump_Pminusvaries} shows two of the relevant fixed point solution branches that are qualitatively comparable to the simulation data. The initial (red) and final (green) steady states correspond to the flipping of $s_x$ shown in Fig. \ref{switch_inh}(d) before and after the pulse was added. 
These results are ready to be tested in experiments, while the value of pump power might be given by a spatially averaged one depending on the specific shape of the pumping laser.

\begin{figure}
\includegraphics[width=\columnwidth ]{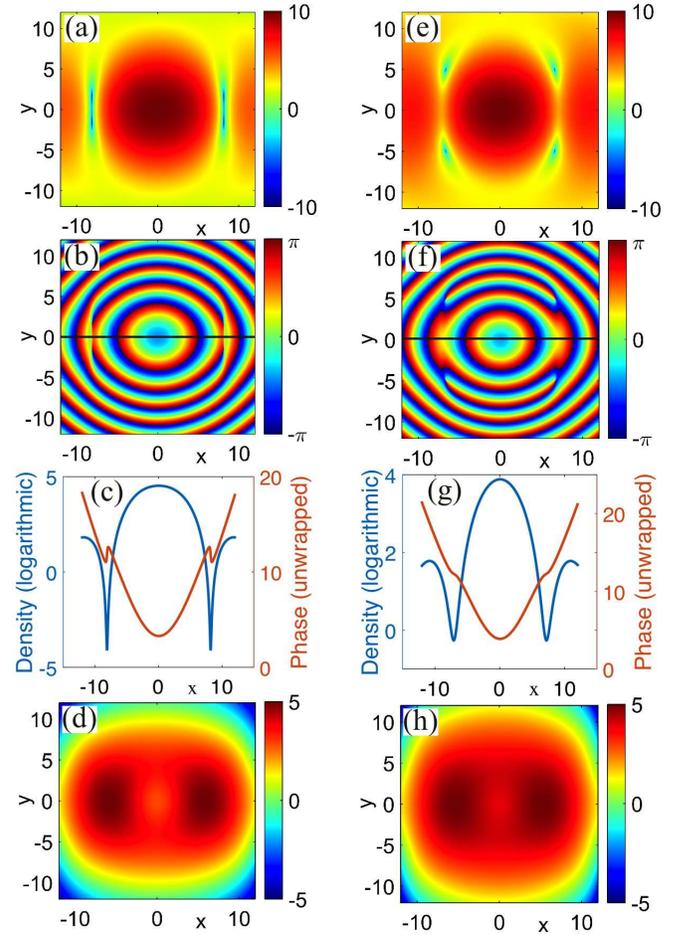}
\caption{(Color online). Spin wave excitation for the case of an elongated pump with the imbalance $P^L_{im}=0.4$ and a tightly focused radially symmetric pulse with $P^L_{im}=0.65$ and different widths (see text).  (a) Density profile of the $\psi_+$ component showing a density dip associated with a phase fold (b). (c) Cross-section of density and phase across the fold. (d) Density profile of the $\psi_-$ component corresponding to (a); (e) Density profile of the $\psi_+$ component showing stationary configuration of vortex pairs connected by a phase fold (f);  (g) Cross-section of density and phase across the phase fold. (h) Density profile of the $\psi_-$ component corresponding to (e). All density plots are logarithmic.}
\label{elongated}
\end{figure}

\section{Excitation of non-trivial spin textures}

While the density and spin wave excitation in the regime of a radially-symmetric {\em cw} pump and pulse leads to the efficient switching between different spin and density states, the spatial dynamics can be captured by the homogeneous or spatial averaging approximation, as long as the radial symmetry of the spatial density distributions for both spin components is preserved. However, the situation changes dramatically when the radial symmetry of the system is broken. This can be achieved by combining a radially symmetric normal incidence pump with elongated excitation, {\em e.g.}, at a steep angle to the sample surface, or vice versa. Fig.~\ref{elongated} demonstrates the consequence of pulsed perturbation of a steady state established with an elongated ($a_x=16$, $a_y=5$) pump. The pulse is a strong, off-resonant, tightly focused Gaussian beam with $a_{px}=a_{py}=4$ (a-d) or $a_{px}=a_{py}=3$ (e-h) and peak power comparable to that of the {\em cw} pump. It was set to having $\pi/2$ phase difference with the pump and thus no interference pattern appeared.
Remarkably, the symmetry of the density flows caused by the pulse is now broken, leading to different flow speeds along the two symmetry axes. As a result, larger area pulse produces a phase fold associated with a pronounced dip in the condensate density Fig. \ref{elongated} (a-c). An even tighter pulse will result in the two phase folds terminating on vortices, leading to formation of a {\em stable and stationary} configuration of two vortex-antivortex pairs.

Although nontrivial phase structures appear only in one of the spin components [see Figs. \ref{elongated} (d), (h)], they are maintained and stabilised by internal Josephson current between the spin components and do not exist in the absence of linear coupling ($J=0$). We note that no phase singularities appear in the linearly-polarized basis, which sets these structures apart from the previously described half-charge vortices \cite{Yuri07,Lagoudakis09,Manni12}

\section{Conclusions}
In conclusion, we have investigated in detail the formation and polarization structure of the steady states of an exciton-polariton condensate created by far off-resonant {\em spin-polarized} pump in a semiconductor microcavity with linear polarization splitting. We model the dynamics of the system under the assumption that the incoherent pump creates a fast-responding and rapidly decaying reservoir, which significantly affects the condensate dynamics. Our analysis includes effectively spatially homogeneous pumping conditions, as well as spatially inhomogeneous pumping with a Gaussian profile. The polarization dynamics in the latter case can be mapped to the homogeneous dynamics through a spatially averaging procedure. We have demonstrated that the phase-locking conditions for existence of stationary states naturally leads to the self-trapped states with a nontrivial phase relationship between the condensate components, including $\pi$ out-of-phase ('anti-bonding') states. The existence of both phase-locked and spin-beating states can enable efficient control and switching of the polarization states in the system. Experimentally, these findings can be tested with spatially averaged polarization measurements under incoherent spin-selective excitation conditions.

In addition, we have analysed formation of non-trivial spin-textures under pulsed, spatially inhomogeneous excitation, and demonstrated numerically reliable creation of stable vortex-antivortex pairs.

Our analysis, with suitable modification of parameters, is widely applicable to a general two-state polariton system with weak linear coupling, including weakly linked spatially separated condensates or multi-mode condensates in shallow potential traps.

\section{ACKNOWLEDGEMENTS}
This work was supported by the Australian Research Council (ARC). O.A.E. acknowledges financial support by the Deutsche Forschungsgemeinschaft (DFG project EG344/2-1) and the Thuringian Ministry for Education, Science and Culture (TMESC project B514-11027). G.L. acknowledges the support of the China Scholarship Council (CSC).


\begin{thebibliography}{99}
\bibitem{parameters} In the numerical calculations, the following physical values of the parameters are chosen as experimentally accessible \cite{BerloffPRB10,Roumpos11,Jvalue}:
$m_{LP}=10^{-4}\:m_e$ (where $m_e$ is the free electron mass), $u_a=6 \times 10^{-3}\:\text{meV}\: \mu \text{m}^2$, $u_b=-0.1u_a$, $g_R=2u_a$, 
$\gamma_c = 0.33 \:\text{ps}^{-1}$, $\gamma_R=1.5\gamma_c$, $R=0.01\: \mu \text{m}^2\: \text{ps}^{-1}$, and $J \approx 0.1\!\sim\! 0.2\: \text{meV}$.
So, the dimensionless quantities are: $u_a=7.7 \times 10^{-3}$, $u_b=-7.7 \times 10^{-4}$, $g_R=1.5\times 10^{-2}$, $\gamma_R=1.5$, $R=8.4\times 10^{-3}$,
and $J\approx 0.1\!\sim\! 1$. This choice of parameters corresponds to an assumption of a very short lifetime of reservoir excitons, however the results do not depend on a particular value of $\gamma_R$.
\end{thebibliography}
\end{document}